\documentstyle[epsbox]{elsart}
\begin{document}
\begin{frontmatter}
 \title{    
   Collapsing open isotropic universe  generated by nonminimally coupled scalar field.
   }   
   \author{ Takahiro Morishima\thanksref{EMAIL-1} }
   \author{ Toshifumi Futamase\thanksref{EMAIL-2} }
   \address{
            Astronomical Institute, Graduate School of Science,  \\
            Tohoku University, Sendai, {\sc 980--8578}, {\sc Japan}
            }
   \thanks[EMAIL-1]{Electronic mail address:  moris@astr.tohoku.ac.jp}
   \thanks[EMAIL-2]{Electronic mail address:  tof@astr.tohoku.ac.jp,\\ \,\,\,\,
                                        Telephone number:  +81-22-217-6512,\,\,\, 
                                        Fax       number:  +81-22-217-6513 } 
\begin{abstract}
We investigated the behavior of an open isotropic universe 
generated by a scalar field which couples with background curvature
nonminimally  
with the coupling constant $\xi$.
In particular we focus on the situation where 
the initial value for the scalar field $\phi_{\rm in}$ is greater than 
the critical value ${\hat \phi}_c$=${m_p}/{\sqrt{8\pi\xi(1-6\xi)}}$.  
The behavior is similar to an open de Sitter universe with $k=-1$ 
with a negative cosmological constant $\Lambda <0$. 
It is found that the universe will collapse eventually to a singularity
and thus has a finite extent in time in the future.  
Furthermore, there are some cases which shows a rebouncing behavior 
before the final collapse. 

%{keywords: }\\
\end{abstract}
 % \pacs{PACS numbers: 98.80.Cq, 98.80.Hw}
 % \begin{keyword}
 %   Cosmology
 % \end{keyword}
\end{frontmatter}

%\narrowtext
%\widetext
%\twocolumn
%\baselineskip 22pt

\section{INTRODUCTION}

There is a growing interest  in the scalar field in the cosmological
situations because of its important role played in the inflationary 
universe scenario\cite{Kolb1973,Linde1991}. 
There the expansion of the universe is totally 
governed by the behavior of the scalar field. 
In this paper we would like to make a comment on the 
behavior of the universe dominated by the scalar field 
from the point of view so far not paid much attention.
Namely  we shall be interested in the universe  
with nonminimally coupled scalar field $\phi$ in some special
circumstances which will be explained in short. 
The nonminimal coupling is described by the form $\frac{1}{2}\xi
R\phi^2$ in the lagrangian
where $R$ is the spacetime curvature and $\xi$ the coupling constant. 
Minimal coupling has $\xi=0$. 
We choose the convention that the conformal invariance 
yields $\xi=\frac{1}{6}$. 
Particle physics do not specify any particular value for 
the coupling constant $\xi$, 
so there is no a priori reason to restrict our attention to a
particular value for the coupling.

There have been many studies on the effect of this coupling 
on the expansion behaviour and on  the chaotic inflationary scenario.
It has been shown that the original chaotic senario\cite{Linde1983} 
does work only for
a limited range of the coupling constant such as $|\xi|< 10^{-3}$
\cite{tof1989a}.  
Fakir and Unruh\cite{Fakir1990a} showed a possibility to have a successiful 
senario of chaotic inflation with a large negative coupling constant.
On the other hand Futamase and Maeda\cite{tof1989a} have pointed out that 
there will be two critical values $\phi_c={mp}/{\sqrt{8\pi\xi}}$ and
 $\hat{\phi}_c={mp}/{\sqrt{8\pi\xi(1-6\xi)}}$ for the scalar field in  
the range $0< \xi < 1/6$ and there is no isotropic solution with flat
and closed spatial curvature if the scalar field is larger 
than $\hat{\phi}_c$. 
Moreover Starobinskiy\cite{Starobinsky1981} 
and Futamase et al\cite{tof1989a,tof1989b}  
have shown that the anisotropic shear diverges as the scalar field 
approaches to  $\phi_c={mp}/{\sqrt{8\pi\xi}}$ for almost all initial 
configurations $\phi > \phi_c$ and thus the closed as well as flat 
universe starting from such initial configurations does
not lead to our present isotropic universe. 
Therefore no serious  attention has been paid for the
situation with such initial configurations for the scalar field. 

However some of the recent observations do suggest the possibility 
of open universe\cite{henry} \cite{SNIa} 
and there are also studies to explore the possibility to
have open inflationary scenario in theoretical side\cite{Gott1982,Sasaki,Bucher1995}. 
In this situation 
we think it may be interesting to consider the case of open universe
with nonminimally coupled scalar field under the condition  $\phi >
\phi_c$ because the possibility to have an isotropic expansion 
in this situations is not excluded and thus there will be no a priori
reason to reject such a situation. 

It turns out that such a universe has remarkably rich behavior 
such that the universe recollapses to a singularity in a finite time.
In this sense the effect of nonminimal coupling constant 
is similar with that of negative cosmological constant. 
More than that there are situations where the  universe show
rebouncing behavior before the final collapse.

This paper is organized as follows. 
In Sec. 2 we give general discussions including a short review  
about the de Sitter universe for reader's convenience and about
nonminimal coupling. 
In Sec. 3 we turn our attention to the case of open universe
($k=-1$). 
There we will find some very interesting behavior of such a universe.
Finally summary and remarks are give in Sec. 4.

\section{General consideration}
\subsection{de Sitter universe}
We briefly review the expansion behavior of de Sitter universe. 
In this model the cosmological constant is assumed to be 
$\Lambda \ne 0$ without matter. 
The space-time is described to be 
the Friedmann-Robertson-Walker (FRW) type: 
\begin{equation}
  ds^2 =  -dt^2 + a^2(t) \, d\Omega_3^2(k)
  \label{eq: metric}
\end{equation}
where $d\Omega_3^2(k)$ is the metric of the universe depending on the
curvature constant $k=+1$, 0 or $-1$, respectively. 
The Einstein equations is  
\begin{equation}
     H^2 + \frac{k}{a^2} = \frac{\Lambda}{3}
    \label{eq: de-sitter}
\end{equation} 
where $H=\frac{\dot{a}}{a}$ is the Hubble expansion rate.

  \begin{figure}         
  \centering
  \psbox[scale=0.5]{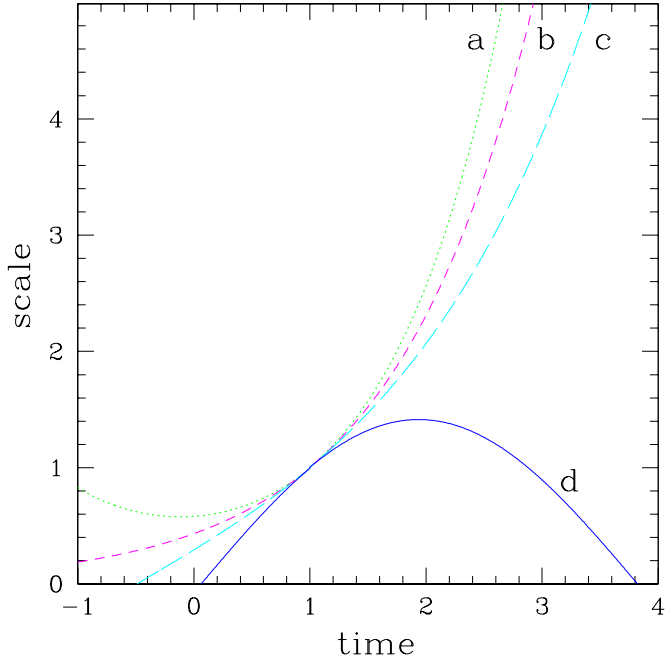}
 %\vspace*{40mm}
  \caption{ 
             The typical behavior of cosmological scale-factor in 
             the de Sitter universe. 
          }
  \label{fig: FIG_deSitter}
  \end{figure}
Fig. \ref{fig: FIG_deSitter},  
this diagram shows the typical behavior of cosmological 
scale-factor in the de Sitter universe. 
Those behavior are apparently different in the case 
of $\Lambda>0$ and $\Lambda<0$.  

The lines (labeled $a$, $b$, $c$) are curves 
in the case of cosmological constant $\Lambda>0$  
with curvature $k>0$, $k=0$ and $k<0$, respectively. 
In the inflationary scenario, the vacuum energy of the 
scalar field (or inflaton) $\phi$ plays the same role with 
the cosmological constant $\Lambda(>0)$.

The solid line (the lowest line labeled $d$) is curve 
in the case of $\Lambda<0$ with $k<0$,  
which is the only case of solution in de Sitter universe
with $\Lambda<0$. 
Here the universe necessarily recollapses and thus 
the spacetime has a finite extent in time in the future. 
Those are limited by the value of cosmological constant $|\Lambda|$. 
Though there are some interesting features in this case, 
it is usually regarded as no physical meaning.

\subsection{Non-minimal coupling constant $\xi$}
We consider the total action of Einstein gravity and a real scalar
field $\phi$ coupled non-minimally with the spacetime curvature.   
\begin{equation}
 S= \int dx^4 \sqrt{-g} 
         \left[ \frac{1}{2}R - \frac{1}{2}(\nabla\phi)^2
	          - V(\phi) -  \frac{1}{2}\xi R\phi^2 \right],
  \label{eq: action}
\end{equation}
where $\xi$ is the coupling constant between the scalar field $\phi$ 
and the space-time curvature $R$. $\xi=0$ and $\xi={1}/{6}$ 
correspond to minimal and conformal couplings, respectively. 
$V(\phi)$ is the potential for the scalar field.

From the above expression, we find that an effective gravitational
constant 
\begin{equation}
 {G_{\rm eff}}/{G}=({1-{\phi^2}/{\phi_c^2}})^{-1} ,
\end{equation}
where 
\begin{equation}
 \phi_c \equiv \frac{m_p}{\sqrt{8\pi\xi}}.
\end{equation}
Thus it  gives us a negative effective gravitational
constant for a reasonable choice of $\xi > 0$, i.e., 
$G_{\rm eff}/G < 0$ for $\phi>\phi_c={m_p}/{\sqrt{8\pi\xi}}$. 
As mentioned in the introduction it has been shown that 
the anisotropic shear diverges as $\phi$ approaches to $\phi_c$ 
in the cases of flat and closed 
geometry\cite{tof1989a,tof1989b,Starobinsky1981}.

When the coupling constant is in the range $0<\xi<\frac{1}{6}$, there 
is another singular point 
\begin{equation}
 \hat{\phi}_c \equiv \frac{m_p}{\sqrt{8\pi\xi(1-6\xi)}}
\end{equation}
This may be seen from the structure of the Hamiltonian constrain\cite{tof1989a}.
\begin{equation}
  {\cal H} =  \left( H - \frac{{\phi\dot{\phi}}/{\phi_c^2}}{1 - {\phi^2}/{\phi_c^2}} \right)^2 
             + \frac{k}{a^2}
             - \left(  \frac{4\pi\dot{\phi}^2}{3} \frac{(1-{{\phi}^2}/{\hat{\phi}_c^2})}{(1-{\phi^2}/{\phi_c^2})^2} 
                     + \frac{8\pi}{3} \frac{V(\phi)}{(1-{\phi^2}/{\phi_c^2})}  \right) 
  \label{eq: Hamitonian}
\end{equation} 
The constraint also indicates that there is no isotropic classical
solution for $\phi >  \hat{\phi}_c$ in the closed or flat universe. 

Only possibility to have an isotropic spacetime for $\phi >
\hat{\phi}_c$ is the case where the
spatial curvature is negative $k = -1$ which we will consider in
detail in the following section.

\section{$k=-1$ open universe with non-minimal coupling}
We now consider an open isotropic universe in the presence of a
nonminimally coupled scalar field. 
When we restrict our consideration in FRW spacetime 
denoted by eq. (\ref{eq: metric}),  
the Einstein equations are found to be 
\begin{equation}
   \left( 1 -  \frac{\phi^2}{\phi_c^2} \right)  
        \left[   \dot{\alpha}^2  + k\,e^{-2\alpha}  \right] 
 = \frac{8\pi}{3} \left[ \frac{\dot{\phi}^2}{2} + 6\xi\dot{\alpha}\dot{\phi}\phi + V(\phi) \right], 
   \label{eq: einstein}
\end{equation} 
where $\alpha=\ln a(t)$ 
and dot denotes the derivative with respect to time. 
The scalar field equation is found to be 
\begin{equation}
   \ddot{\phi} + 3\dot{\alpha}\dot{\phi} + V'_{\rm eff}(\phi) = 0
   \label{eq: phi-field}
\end{equation} 
where $ V_{\rm eff}(\phi) $ is effective potential and its 
gradient is written as  
\begin{eqnarray}
   V'_{\rm eff}(\phi) &=& V'(\phi) + \xi R  \\
            &=& (1-\frac{\phi^2}{{\hat \phi}_c^2})^{-1}
                \left[
                - \frac{\phi{\dot \phi}^2}{{\hat \phi}_c^2} + (1-\frac{\phi^2}{\phi_c^2}) V'(\phi)
                + \frac{4\phi V(\phi) }{\phi_c^2}
                \right]  
\end{eqnarray} 
In the following we assume $V(\phi)=\lambda\phi^4/4!$ 
with $\lambda=0.01$ as a typical example.
We shall then numerically solve the above set of 
equations (\ref{eq: einstein}) and (\ref{eq: phi-field}) 
for the case $0<\phi_c<\phi_{\rm in}$. 
We present the results of calculation 
with $\xi=0.1$(${\hat \phi}_c \simeq 1$), $\dot{\phi}_{\rm in}=-1.0$
and $\phi_{\rm in}=5, 10, 15, 20, 25$.
We notice that there is upper limit for the initial value 
about 30 in this case otherwise one cannot satisfy 
eq. (\ref{eq: einstein}) at the initial time.

  \begin{figure}         
   \psbox[scale=0.4]{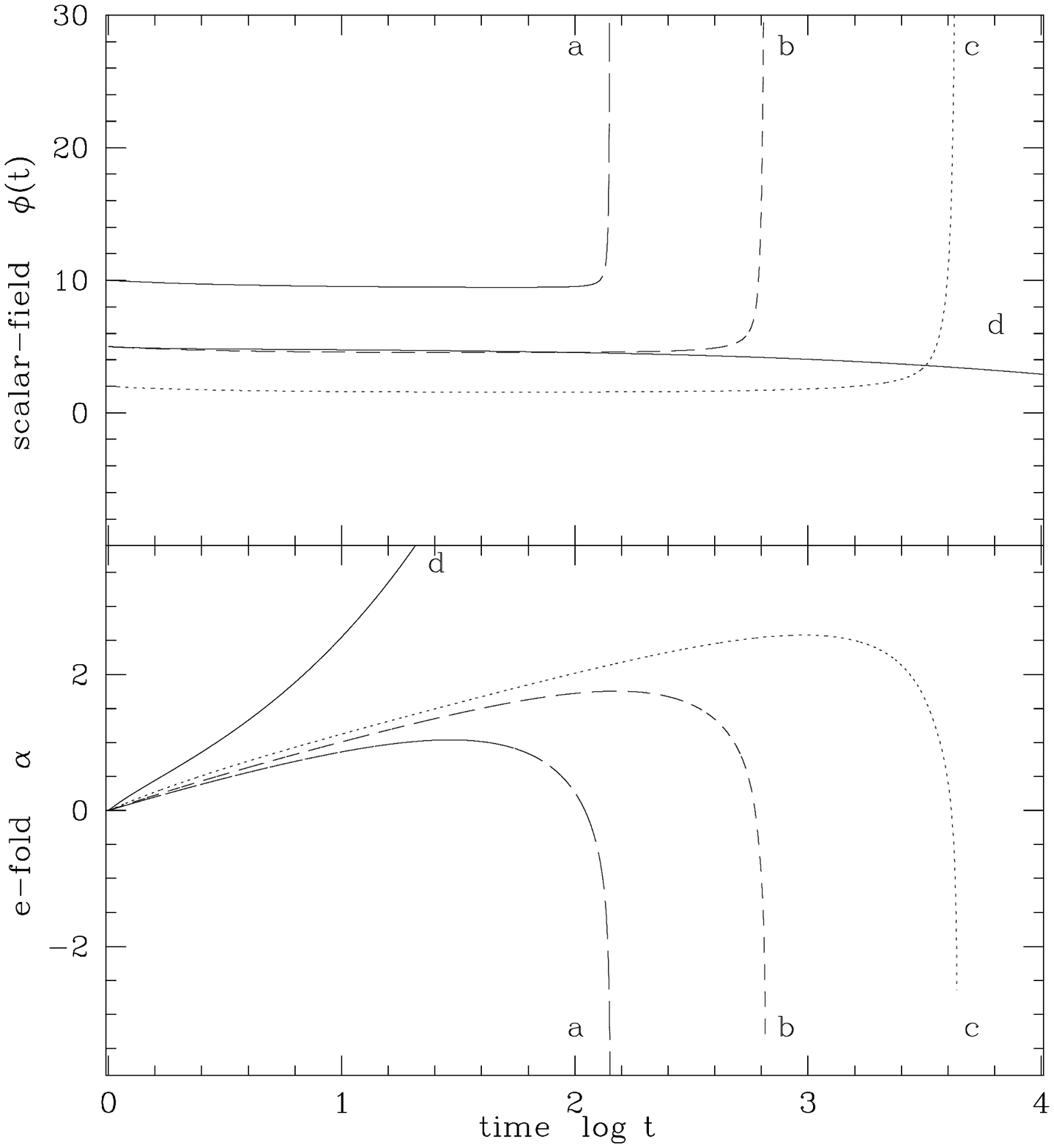}
   \centering
   %\vspace*{40mm}
    \caption{ 
              The evolutions of the open isotropic universe 
              generated by nonminimally coupled scalar field. 
              The upper panel describes evolution of scalar field $\phi$ 
              and the lower panel describes that of cosmological scale factor 
              with the initial value of scalar field as $\phi_{\rm in}$ =10, 5, 2,
            }
    \label{fig: FIG_2}
    \psbox[scale=0.4]{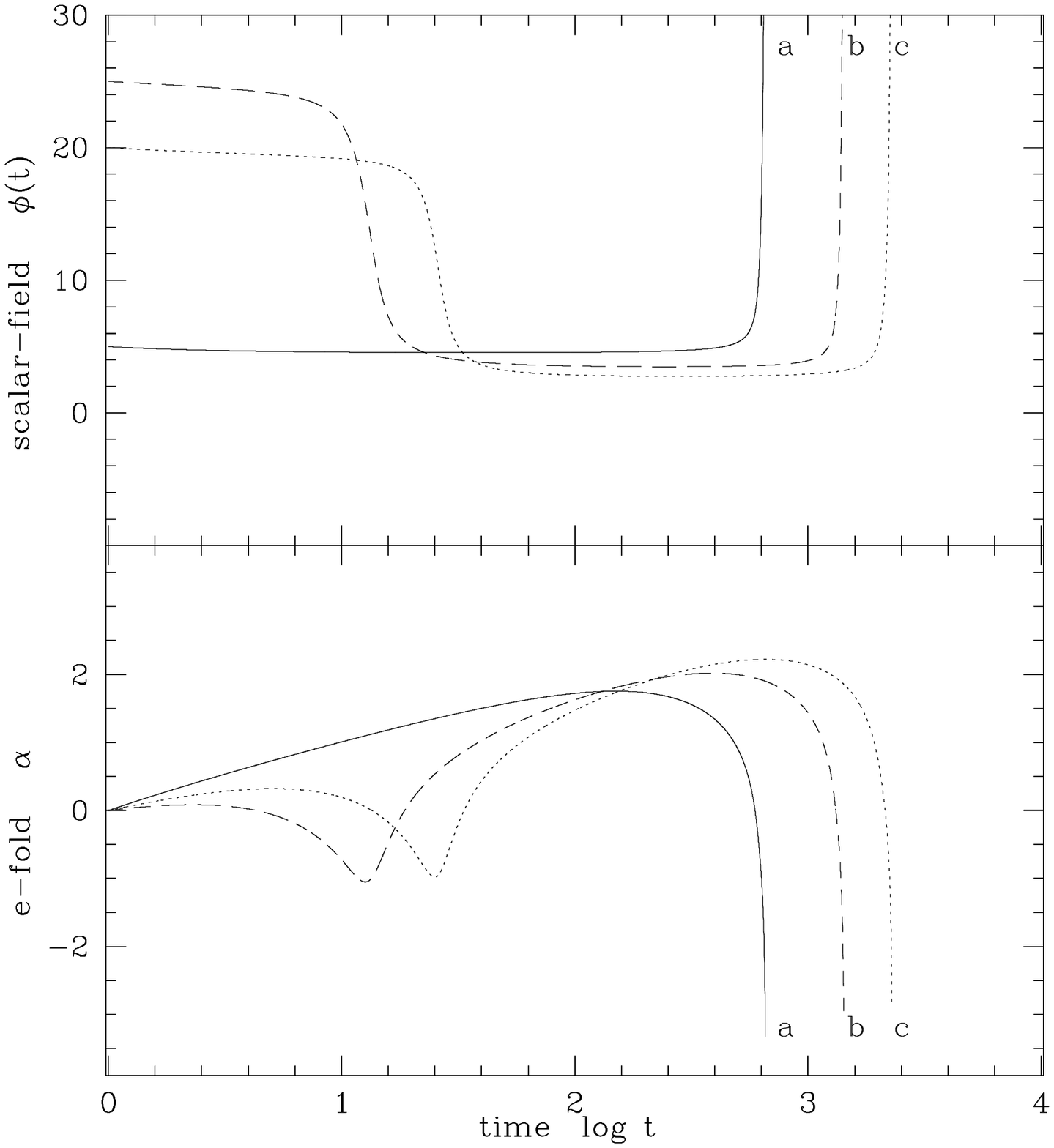}
    \centering
    %\vspace*{40mm}
    \caption{ 
              As Fig. 2, but for $\phi_{\rm in}$ =5, 20, 25.
            }
    \label{fig: FIG_3}
  \end{figure}
Fig.\ref{fig: FIG_2} shows evolutions of the scalar field (upper
figure) and the scale factor( lower figure). 
Those are scaled in Planck unit.  
The lines (labeled $a$, $b$, $c$) are solutions 
with the initial value of scalar field 
as $\phi_{\rm in}$ =10, 5, 2, respectively. 
These $k=-1$ open universe look like $k=-1$ open  
anti-de Sitter universe with negative cosmological constant 
$\Lambda<0$ in Fig.\ref{fig: FIG_deSitter}.  
We also draw the solid curve (labeled $d$) 
in the case of $\xi=10^{-11}$ with $\phi_{\rm in}$ = 5 under $k$=+1 
closed universe for comparison. 
This is a most popular case known as 
{\it chaotic inflationary universe}\cite{Linde1983,Linde1991}.

Fig.\ref{fig: FIG_3} is the same as Fig. \ref{fig: FIG_2}, 
but the lines (labeled $a$, $b$, $c$) are solutions with 
the initial value of scalar field as 
$\phi_{\rm in}$ =5, 20, 25, respectively. 
In the case of $\phi_{\rm in}$ =5, 
the scalar field $\phi$ seems to be blocked by barrier of the critical
value ${\hat \phi}_c \simeq 1$ so as not to go through it  
for a long time, and then  the scalar field $\phi$ blows up to
infinitely large value. This behavior may be understood by noticing 
that the gradient of the effective potential diverges as the scalar
field approaches to ${\hat \phi}_c$.          
The evolution of scale factor looks like that of de Sitter
universe with $k<0$ and $\Lambda<0$, which necessarily 
come to collapse into zero ($\alpha=-\infty$) in a finite time.
In the case of large initial values such as $\phi_{\rm in} > 11$,
the scalar field $\phi$ decreases first to near the  ${\hat \phi}_c$,  
keep a value for a while, and blow up go to infinitely large value. 
Then the evolution of scale factor experiences a bounce era  
and goes to recollapse in the end.

\section{Conclusion}
%\section{DISCUSSION}
We investigated the evolutionary behavior of an open isotropic 
universe dominate by a nonminimally coupled scalar field in the range 
 $0<{\hat \phi}_c<\phi_{\rm in}$. We found by numerical analysis that 
the scalar field cannot cross the singular point ${\hat \phi}_c$ 
and instead diverges to infinity. According to the behavior of the
scalar field, the expansion of the universe is turn around and
is collapsing to singularity.  This behavior is similar to 
an open de Sitter universe with a 
negative cosmological constant eq. (\ref{eq: de-sitter}).
This could be understood by writing the expansion equation in the
following form.
\begin{equation}
   \dot{\alpha}^2 = - \frac{8\pi}{3} 
                      \frac{\left[ {\dot{\phi}^2}/{2} 
                                 + 6\xi\dot{\alpha}\dot{\phi}\phi 
                                 + V(\phi) \right]}
                           {({\phi^2}/{\phi_c^2} -1)}
                    -  k \,\,e^{-2\alpha} 
\end{equation} 
Thus the energy density of the scalar field plays a role of a negative
cosmological constant in a sense.  
If the cosmological constant $\lambda_0$ is not zero, 
the effective cosmological constant $\lambda_{\rm effect}$ 
is found to be 
\begin{equation}
  \lambda_{\rm effect} (\phi, \xi) = \frac{\lambda_0}{(1-\phi^2/\phi_c^2)}
                 + \frac{8\pi}{3}\frac{({\dot\phi}
^2/2+6\xi{\dot \phi}\phi+V(\phi))}{(1-\phi^2/\phi_c^2)}.
\end{equation}
It is found that non positive effective cosmological constant 
$\lambda_{\rm effect}$ appears  
even if $\lambda_0 >0$ in some range of the scalar field 
$\phi_{\rm in}> \phi_c$. Thus it might gives us a possible mechanism
to reduce the cosmological constant to a small value.

There are also other solutions which shows a rebouncing behavior  
before the eventual collapse to the singularity. 
It seems to us that this kind of behavior has not been known before. 
Just after the bounce the expansion is accelerated. 
Then it might be possible to have a new open
inflationary scenario by adjusting parameters (initial value of the
scalar field and the coupling constant).

It also interesting to consider how anisotropic shear behaves in the
singular point in our open universe. Our preliminary result show that 
the anisotropy diverges also as one approaches to $\phi_c$ also in the
open universe. This will be published elsewhere.

%\section*{acknowledgments}
%We wish to thank .................

%\narrowtext
%\widetext
%\mediumtext

\end{document}